\documentclass[preprint,a4paper]{elsarticle}
\usepackage{amsmath,amsfonts,euscript,amssymb}
\usepackage{epsfig}
\usepackage{graphicx}

\newcommand{\be}{\begin{equation}}
\newcommand{\ee}{\end{equation}}
\newcommand{\bea}{\begin{eqnarray}}
\newcommand{\eea}{\end{eqnarray}}

\begin{document}

\begin{frontmatter}

\title{Conformal Hamiltonian  Dynamics of General Relativity}

\author[bltp]{A.B.~Arbuzov}
\author[bltp]{B.M.~Barbashov}
\author[bltp,pdm]{R.G.~Nazmitdinov}
\author[bltp]{V.N.~Pervushin\corref{cor1}}
\ead{pervush@theor.jinr.ru}
\author[born]{A.~Borowiec}
\author[kir]{K.N.~Pichugin}
\author[itep]{A.F.~Zakharov}

\cortext[cor1]{Corresponding author}

\address[bltp]{Bogoliubov Laboratory of Theoretical Physics,
Joint Institute for Nuclear Research, 141980 Dubna, Russia}
\address[pdm]{Department de F{\'\i}sica,
Universitat de les Illes Balears, E-07122 Palma de Mallorca, Spain}
\address[born]{Institute of Theoretical Physics, University of
 Wroc\l aw, Pl. Maxa Borna 9, 50-204 Wroc\l aw, Poland}
\address[kir]{Kirensky Institute of Physics, 660036 Krasnoyarsk,
 Russia}
\address[itep]{Institute of Theoretical and
Experimental Physics, B. Cheremushkinskaya str. 25, 117259 Moscow, 
Russia}

\begin{abstract}
The General Relativity  formulated  with the aid of the spin 
connection coefficients is considered  in the finite space geometry 
of similarity with the Dirac scalar dilaton. We show that the 
redshift  evolution of the General Relativity describes the vacuum 
creation of the matter in the empty Universe at the electroweak 
epoch and the dilaton vacuum energy plays a role of the dark energy.
\end{abstract}


\end{frontmatter}

Submitted to Phys. Lett. B 14 April 2010;  accepted 23 June 2002.

\vspace{1cm}

 The distance-redshift dependence in the  data of the 
type Ia supernovae \cite{SN2} is a topical problem in  the standard 
cosmology (SC). As it is known, the SNeIa  distances are greater 
than the ones predicted by the SC  based on the matter dominance  
idea \cite{eds32}.  There are numerous attempts to resolve this 
problem with a various degree of success (see for review 
\cite{Linde:2007fr}). One of the popular approaches is the  
$\Lambda$-Cold-Dark-Matter model \cite{gio}. It provides, however, 
the present-day slow inflation density that is less by factor of 
$10^{-57}$  than the  fast primordial inflation density  proposed to 
include  the Planck epoch.  

Approaches to the General Relativity (GR) with conformal symmetry   
provide a natural relation to the SC \cite{Grum}. 
The Dirac  version \cite{Dirac_73} of the geometry of similarity 
\cite{Weyl_18} is an efficient way to include the conformal symmetry into the GR.
In fact, the latter approach allows to
explain the SNeIa data without the inflation \cite{Behnke_02}. In 
the present paper, the Dirac formulation of the GR in the geometry of 
similarity is adapted to the diffeo-invariant Hamiltonian approach  
by means of the spin connection coefficients in a finite space-time, 
developed in \cite{Barbashov_06}. In this way we study a possibility 
to choose variables and their initial data that are compatible with 
the observational data  associated with the dark energy content.  We 
find integrals of motion of the metric and matter fields in terms of 
the variables distinguished by the conformal initial data.

Within the Dirac  approach the Einstein-Hilbert action takes the 
form
 \bea 
 &W_{\rm Hilbert}=
 -\int\limits_{ }^{ }d^4x|-e|
 \frac{1}{6}R(e)\bigg |_{e=e^{-D}\widetilde{e}} \nonumber\\
 & = -  \int\limits_{ }^{ } d^4x 
 \biggl[\frac{|-\widetilde{e}|e^{-2D}}{6}R(\widetilde{e})-e^{-D}\partial_\mu
 \left(|-\widetilde{e}|\widetilde{g}^{\mu\nu}\partial_\nu e^{-D}\right)\biggr] .
 \label{sGR2} 
 \eea 
Hereafter, we use the units $M^2_{\rm Planck}\frac{3}{8\pi}=1$. The 
interval is defined via diffeo-invariant linear forms 
$\omega_{(\alpha)}=e_{(\lambda) \mu}dx^\mu$ with the  tetrad 
coefficients 
 \be\label{1ph} ds^2\!=\!g_{\mu\nu}dx^\mu
dx^\nu=\omega_{({\alpha}) 
}(d)\otimes\omega_{({\beta})
 }(d)\eta_{({\alpha})({\beta})
};~~ \eta_{({\alpha})({\beta})
}\!=\!Diag (1,\!-1,\!-1,\!-1).
 \ee
 The geometry of similarity \cite{Dirac_73,Weyl_18} means  the 
identification of measured physical quantities ${F}^{(n)}$, where 
$(n)$ is the conformal weight, with their ratios  in dilaton units 
$e^{-nD}$ 
 \be\label{ratio-1} 
 \widetilde{F}^{(n)}=e^{nD}{F}^{(n)},\qquad \widetilde{ds}^2=e^{2D} ds^2.
 \ee 
 We define the measurable space-time coordinates in the  GR 
as the scale-invariant quantities in the framework of the Dirac-ADM 
4=1+3 foliation \cite{dir,ADM}  
 \bea \label{dg-2a} 
\widetilde{ds}^2=
 {\widetilde{\omega}}_{(0)}\otimes{\widetilde{\omega}}_{(0)}-
{\widetilde{\omega}}_{(b)}\otimes{\widetilde{\omega}}_{(b)}, 
 \eea
 where the linear forms ${\widetilde{\omega}}_{(\alpha)}=\widetilde{e}_{(\alpha)\mu}dx^\mu$ are
  \bea \label{dg-2aa}
&&{\widetilde{\omega}}_{(0)}=e^{-2{D}}N dx^0,\\
\label{dg-3a} 
&&{\widetilde{\omega}}_{(b)}=\overline{\omega}_{(b)}+ {\bf e}_{(b)j}N^jdx^0,
\\
\label{1-2c5} &&{\overline{\omega}}_{(b)}={\bf e}_{(b)i}dx^i.
 \eea
Here ${N}$ is the Dirac lapse function, $N^j$ are the shift vector 
components, and ${\bf e}_{(b)i}$ are the triads corresponding to the 
unit spatial metric determinant 
$|\widetilde{g}^{(3)}_{ij}|\equiv|{\bf e}_{(b)j}{\bf e}_{(b)i}|=1$. 

The Dirac dilaton $D= -(1/6)\log|{g}^{(3)}_{ij}|=\langle D\rangle+{\overline{D}}$,
is taken in the Lichnerowicz gauge~\cite{lich}.
The Dirac lapse function 
${N}=N_0(x^0){\cal N}(\tau,x)$ is split on the global factor 
$N_0^{-1}=\langle N^{-1}\rangle$ which determines all time intervals 
used in the  observational cosmology: the redshift interval 
$d\tau=N_0dx^0$ \cite{Misner_69}, 
the conformal one $d\eta=d\tau e^{-2\langle D\rangle}$, and the world 
interval $dt=e^{-\langle D\rangle}d\eta=d\tau e^{-3\langle D\rangle}$.
In this case the dilaton zeroth mode $\langle D\rangle=V_0^{-1}\int_{V_0} d^3x D$ 
(defined in the finite diffeo-invariant volume) coincides with the logarithm of
the redshift of spectral line energy $E_{\rm m}$
\be
\label{red-1}
\langle D\rangle=\log(1+z)=\log \left( E_{\rm m}(\eta_0-\eta)/E_{\rm m}(\eta_0)\right),
\ee
where $\eta_0$ is the present-day conformal time interval, and 
$\eta_0-\eta=r/c$ is the SNeIa distance. 
In accord with the new Poincar\'e group classification, the "redshift" 
(\ref{red-1}) is treated as one of the 
matter components, on the equal footing with the matter.  
  
 The key point of our approach is to express the GR action 
directly in terms of the redshift factor. The action can be 
represented as a sum of the dilaton and the graviton terms:
 \bea
\label{1-3n}
 &{W}_{\rm Hilbert}=
\int\limits_{}^{}
 dx^0\left[-\dfrac{\left(\partial_0\langle D\rangle\right)^2}{N_0}
 +N_0e^{-2\langle D\rangle}\textsf{L}_g\right],\\
 \label{1-3nL}
 &\textsf{L}_g=e^{2\langle D\rangle}\int d^3x {\cal N}
\left[-(v_{\overline{D}})^2+\dfrac{v^2_{(ab)}}{6}-e^{-4D}\dfrac{R^{(3)}}{6}\right].
 \eea
 Here, 
 \bea 
\label{1-17a}
 R^{(3)}&=&R^{(3)}({\bf e})-
 \frac{4}{3}{e^{7D/2}}\triangle e^{-D/2},
\eea
is the curvature, where
$R^{(3)}({\bf e})$ is expressed via the   spin-connection 
coefficients  \bea \label{1-19} 
\omega^{\pm}_{(ab)}(\partial_{(c)})&=&\frac{1}{2}\left[ {\bf 
e}^j_{(a)}\partial_{(c)}{\bf e}^j_{(b)}\pm{\bf
e}^i_{(b)}\partial_{(c)}{\bf e}^i_{(a)}\right]\,\,,
 \eea
and $\triangle =\partial_i[{\bf e}^i_{(a)}{\bf e}^j_{(a)}\partial_j]$ 
is the Laplace operator.  

The dependence of the linear forms 
 \be {\overline{\omega}}_{(b)}(d)={\bf e}_{(b)i}dx^i
 = dX_{(b)}-X_{(c)}{\bf e}^i_{(c)}d{\bf e}_{(b)i} 
 \label{1-2cj1} 
 \ee 
on the tangent space coordinates $X_{(b)}\equiv\int dx^i {\bf 
e}_{(b)i}= x^i {\bf e}_{(b)i}$ by means of the spin connection coefficients 
can be obtained by virtue of the Leibniz rule $AdB=d (AB)-(AB)d\log 
(A)$ (in particular $d[x^i]{\bf e}^{T}_{\underline{b}i}=d[x^i{\bf 
e}^{T}_{\underline{b}i}]-x^id[{\bf e}^{T}_{\underline{b}i}]$). 
The difference between this approach to gravitation waves  
and the accepted one \cite{Gr-74,ps1} is that the symmetry with 
respect to diffeomorphisms is imposed on spin connection 
coefficients. 

The linear graviton form (\ref{1-19}) can be expressed via two 
photon-like polarization vectors 
$\varepsilon^{(\alpha)}_{(a)}(\textbf{k})$. By virtue of the  
condition 
 \bea
{\sum_{\alpha=1,2}\varepsilon^{(\alpha)}_{(a)}(\textbf{k})\varepsilon^{(\alpha)}_{(b)} 
(\textbf{k})=\delta_{(a)(b)}-\dfrac{\textbf{k}_{(a)} 
\textbf{k}_{(b)}}{\textbf{k}^{(2)}}}, 
 \eea
one obtains
  \bea
 \label{R-1}
\omega^+_{(ab)}( \partial_{(c)}) =\sum_{\textbf{k}^2\not =0} 
\dfrac{e^{i\textbf{\textbf{k}\textbf{X}}}}{\sqrt{2\omega_{\textbf{k}}}} 
\textbf{k}_{(c)}[\varepsilon^R_{(ab)}
(\textbf{k}) g_{\textbf{k}}^+(\eta)
+\varepsilon^R_{(ab)} (-\textbf{k})g_{\textbf{k}}^-(\eta)],
 \eea
 where  
 ${\varepsilon}^R_{(ab)}(\textbf{k})={\rm diag}[1,-1,0]$ in the orthogonal  
 basis of spatial vectors 
$[\vec{\varepsilon}\,^{(1)}(\textbf{k})$, $\vec{\varepsilon}\,^{(2)}(\textbf{k})$, 
${\textbf{k}}]$. 
Here, $g^{\pm}$ are the holomorphic variables of the single degree 
of freedom,  $\omega_{\textbf{k}}=\sqrt{{\textbf{k}}^2}$ is the 
graviton energy normalized (like a photon in QED) on the units 
of a volume and time 
 \be \label{3-1pp}
  \overline{g}_{\textbf{k}}^{\pm}=\dfrac{\sqrt{8\pi}}{M_{\rm
  Planck}V^{1/2}_0}g_{\textbf{k}}^{\pm}. 
 \ee
The triad velocities  
 \bea 
\label{vab-1}
&v_{(ab)}=\frac{1}{N}\left[\omega^+_{(a)(b)}(\partial_0-N^l\partial_l)+
\partial_{(a)}N^{\bot}_{(b)}+\partial_{(a)}N^{\bot}_{(b)}\right]
 \eea 
depend on the symmetric forms $\omega^+_{(ab)}$, and  the shift 
vector components $\partial_{(b)}N^{\bot}_{(b)}=0 $ are treated as 
the non-dynamical potentials. This means that the anti-symmetric
forms  $\omega^-_{(ab)}$ are not dynamically independent variables 
but are determined by a matter distribution. 

Following Dirac \cite{dir,Fad} one can define  such a coordinate 
system, where  the covariant velocity $v_{{\overline{D}}}$ of the 
local volume element and the momentum 
 \be 
 \label{dirac-1}
P_{{\overline{D}}}=2v_{\overline{D}}= 
\frac{2}{N}\left[(\partial_0-N^l\partial_l){\overline{D}} 
+\partial_lN^l/3\right]=0
 \ee 
are zero. As a result, the dilaton deviation   $\overline{D}$ can be 
treated as a static potential.  The dilaton contribution to the 
curvature (\ref{1-17a}) with matter sources yield the Schwarzschild 
solution  of classical equations 
$\triangle[\exp\{-7\overline{D}/2\}{\cal N}]=0$ and 
$\triangle\exp\{-\overline{D}/2\}=0$. The solutions are 
$\exp\{-7\overline{D}/2\}{\cal N}=1+r_g/(4r)$ and 
$\exp\{-\overline{D}/2\}=1-r_g/(4r)$  in the isotropic coordinates 
of the Einstein interval $ds$, where $r_g$ is the gravitation radius 
of a matter source. These solutions double the angle of the photon 
beam deflection by the Sun field, exactly as the Einstein's metric 
determinant.  Note that the  GR theory provides also the Newtonian 
limit in our variables (see details in \cite{Barbashov_06}). 
Furthermore, in empty space without a matter source $(r_g=0)$, the 
mean field approximation (${\cal N}=1$, $\overline{D}=0$, $N^l=0$) 
becomes exact.

If there are no matter sources one can impose the condition 
$\omega^-_{(a)(b)}=0$, since the kinetic term~(\ref{vab-1}) depends 
only on $\omega^+_{(ab)}$ components.  In this case the curvature 
(\ref{1-17a}) takes the bilinear form 
 \be
 \label{cur-1} 
 R^{(3)}({\bf e}) 
 =\omega^+_{(ab)}(\partial_{(c)})\omega^+_{(ab)}(\partial_{(c)}). 
 \ee 

The variation of the Hilbert  action with respect to the lapse 
function leads to the energy constraint \cite{z-06} 
 \bea \label{c-1} 
(\partial_\tau\langle D\rangle)^2&=&\rho_{\rm cr}\Omega_{\langle 
D\rangle}+e^{-2\langle D\rangle}\textsf{H}_g/V_0\,\,, 
 \eea 
where the dilaton integral of  motion $\rho_{\rm cr} \Omega_{\langle 
D\rangle}$ is added,  $\rho_{\rm cr}=H^2_0M^2_{\rm Pl}3/(8\pi)$ is 
the critical density, and
 \bea
 \label{1-3nH}
 &\textsf{H}_g= e^{2\langle D\rangle}\int d^3x {\cal N}\left[3p^2_{(ab)}+\dfrac{e^{-4D}R^{(3)}}{6}
\right]
 \eea
is the graviton Hamiltonian, $p_{(ab)}=v_{(ab)}/3$ is a canonical 
momentum (see Eq.(\ref{vab-1})).

Straightforward calculations define a set of  
 evolution equations for the Lagrangian 
$\textsf{L}_g$ (\ref{1-3nL}) and the Hamiltonian $\textsf{H}_g$ 
(\ref{1-3nH}) 
\bea
\label{3-9}
\partial_{\langle D\rangle} \textsf{H}_g&=&2\textsf{L}_g,\\
\label{3-10}
\partial_{\langle D\rangle} \textsf{T}_g&=&2e^{-2\langle D\rangle}\textsf{L}_g,\\
\label{3-11}
\partial_{\langle D\rangle} \textsf{L}_g&=&2\textsf{H}_g-2e^{-2\langle D\rangle}\textsf{T}_g,
\eea 
where $\textsf{T}_g=\sqrt{\textsf{H}_g^2-\textsf{L}_g^2}$. 

Note, the GR equations in terms of the  spin-connection coefficients  
(\ref{3-9})-(\ref{3-11}) 
 coincide with the evolution equations for the  
 parameters of squeezing $r_b$ and rotation $\theta_b$  \cite{grib88} 
 \bea \label{p-1} 
 \partial_{\langle D\rangle} r_b&=&\cos 2\theta_b,
 \\ \label{p-2} 
 \omega_{\textbf{\rm so}}-\partial_{\langle D\rangle}\theta_b 
 &=& \coth 2r_b\sin 2\theta_b 
 \eea
of the Bogoliubov transformations $A^+=B^+\cosh r 
e^{i\theta}+B^-\sinh r e^{i\theta}$ for a squeezed oscillator (SO) 
$\partial_{\langle D\rangle} 
 A^{\pm}=\pm i \omega_{{\rm so}}A^{\pm}+A^{\mp}$. 
 Indeed, Eqs.(\ref{p-1}),(\ref{p-2}) establish similar 
relations for the expectation values of various combinations of the 
operators $A^{\pm}$ with respect to the Bogoliubov vacuum $B^-|>=0$ 
(see details in \cite{z-06}) 
\bea \label{c-4bn}
 N_b\equiv<|A^+A^-|>&=& \frac{\cosh2r_b-1}{2}\equiv\omega^{-1}_{\textbf{\rm 
so}}:\textsf{H}_{\rm b}:,
 \\  \label{c-4bcc}
\frac{i}{4} <A^-A^--A^+A^+>&=& \frac{\sinh 2r_b\sin 
2\theta_b}{2}\equiv\omega^{-1}_{\textbf{\rm so}}\textsf{T}_{\rm b},
\\ \label{c-4bc} 
 \frac{1}{4} <A^+A^+ + A^-A^->&=&  
\frac{\sinh 2r_b\cos 2\theta_b}{2}\equiv\omega^{-1}_{\textbf{\rm 
so}}\textsf{L}_{\rm b}, 
 \eea

On the other hand, Eqs.~(\ref{1-3nL}), (\ref{R-1}), (\ref{cur-1}), and (\ref{1-3nH}) 
show up that the graviton action~(\ref{1-3n}) has a bilinear oscillator-like form 
\bea
\label{3-8a}
  \textsf{H}_g &=&  \sum\limits_{\textbf{k}}^{}\underline{{\cal H}}_{\textbf{k}},
 \qquad \qquad\qquad \underline{{\cal H}}_{\textbf{k}}
 =\frac{\omega_{\textbf{k}}}{2}
 [g_{\textbf{k}}^+g_{-\textbf{k}}^-+g_{\textbf{k}}^-g_{-\textbf{k}}^+],\nonumber\\
 \textsf{L}_g &=&  \sum\limits_{\textbf{k}}^{}\underline{{\cal L}}_{\textbf{k}},
  \qquad \qquad \qquad \underline{{\cal L}}_{\textbf{k}}=
 \frac{\omega_{\textbf{k}}}{2} [g_{\textbf{k}}^+g_{-\textbf{k}}^++g_{\textbf{k}}^-g_{-\textbf{k}}^-], \\
 \textsf{T}_g &=&  \sum\limits_{\textbf{k}}^{}\underline{{\cal T}}_{\textbf{k}},
  \qquad \qquad \qquad \underline{{\cal T}}_{\textbf{k}}=
\frac{i\omega_{\textbf{k}}}{2} 
[g_{\textbf{k}}^+g_{-\textbf{k}}^+-g_{\textbf{k}}^-g_{-\textbf{k}}^-],\nonumber
 \eea
where 
 \be \label{hv-1} g^{\pm}_{\textbf{k}}=\left[ 
\overline{g}_{\textbf{k}} \sqrt{\omega_{\textbf{k}}} \mp 
ip_{\textbf{k}}/\sqrt{\omega_{\textbf{k}}} \right]/\sqrt{2} 
 \ee 
are the classical variables in the holomorphic representation 
\cite{ps1}. The form (\ref{hv-1}) suggests itself to replace the
variables $g^{\pm}_{\textbf{k}}$ by creation and annihilation graviton operators. 
Evidently, in this case we have to postulate  the existence of a stable vacuum $|0\rangle$.  
As a consequence, it is reasonable to suppose that
the classical graviton Hamiltonian (see Eqs.(\ref{3-8a})) is 
the quantum Hamiltonian averaged over coherent states~\cite{BR}.
One may speculate that such procedure reflects a transformation of a genuine quantum
Hamiltonian (describing the initial dynamics of the Universe) to 
the classical Hamiltonian, associated with  present-day dynamics.

 Having the correspondence
 between two sets of equations (\ref{3-9})-(\ref{3-11}) for the GR and  
(\ref{c-4bn})-(\ref{c-4bc}) for the SO, we are led to the ansatz 
that the SO is the quantum version of our graviton Hamiltonian (see 
also \cite{Gr-74}). This is a central point of our construction. 
As a result, the normal ordering of the graviton Hamiltonian yields
 \be 
\label{3-12g} 
\textsf{H}_g=\textsf{H}_b=:\textsf{H}_b:+\dfrac{\omega_{\textbf{\rm c}}}{2}, 
\qquad
\textsf{L}_g=\textsf{L}_b,
\qquad
 \textsf{T}_g=\textsf{T}_b,
 \ee
 where $\omega_{{\rm c}}=\omega_{\textbf{\rm so}}e^{2\langle D\rangle}$ \cite{z-06}.
The normal ordering creates the Casimir--type vacuum energy 
$\omega_{\textbf{\rm c}}=0.09235/(2r_h)$ \cite{sh-79}, where $r_h$ 
is the radius of the sphere defined by the Hubble parameter. 

The solution of Eqs.~(\ref{3-9})-(\ref{3-11}) is shown at Fig.~1. In 
accordance with this solution, at the tremendous redshift 
$1+z=e^{\langle D\rangle}=a^{-1}$, $z\to \infty$, $a=0$, 
Eq.(\ref{c-1}) is reduced to the zeroth mode dilaton integral of 
motion $\Omega_{\langle D\rangle}$ which corresponds to the 
z-dependence of the Hubble parameter $H(z)=H_0(1+z)^2$. At this 
moment, the Universe was empty, and all particle densities had the 
zero initial data. The same dilaton vacuum regime $H(z)=H_0(1+z)^2$ 
is compatible with the SNeIa data \cite{SN2} in the geometry of 
similarity (\ref{ratio-1}) \cite{Behnke_02}.  

 The next step is the creation of gravitons induced by the direct 
dilaton interaction.  
A hypothetic observer being at the first instance at $r_{\rm I}
=1/H_{\rm I}$ in the primordial volume $V_{\rm I}= 4\pi  r_{\rm 
I}^3/3 
$ observes the vacuum creation of  
these particles with the primordial density
 \bea \label{cr-11} 
 \Omega_{g\rm I}= 
\omega_{\textbf{\rm c}}\cdot \frac{H^2_{0}}{M^2_{\rm Pl.}}\cdot 
(1+z_{\rm I})^{8}
 \eea
defined by the Casimir energy. The question which remains to answer 
is how to define $z_{\rm I}$?

\begin{figure}[t]
 \includegraphics[width=0.55\textwidth,angle=-0]{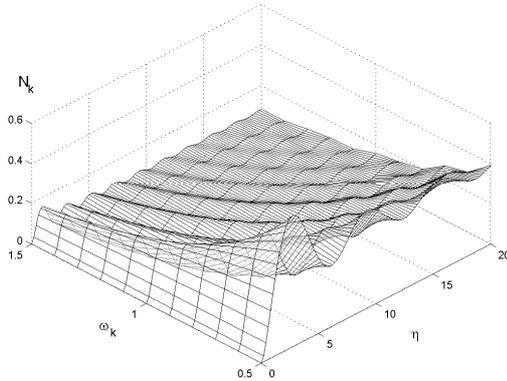}
\parbox{5cm}{
\caption {\label{fig1} The creation of the Universe distribution 
$[N_{\bf k}=N_b]$ (\ref{c-4bn}) versus dimensionless time $\eta$ and 
energies $0.5\leq\omega_{\bf k}$ at the initial data $N_{\bf 
k}(\eta=0)=0$ and the Hubble parameter 
$H(\eta)=1/(1+2\eta)=(1+z)^2$. } }
\end{figure}

In order to estimate the instance of creation $(1+z_{\rm I})$, one 
can add the Hamiltonian of the Standard Model (SM): 
$\textsf{H}_g~\to~ \textsf{H}=\textsf{H}_g+\textsf{H}_{\rm SM}$, -- 
when in the limit $(1+z_{\rm I})\to \infty$ and $a\to 0$ all 
particles become nearly massless $\sqrt{\textbf{k}^2+a^2M^2_0}\to 
\omega_{\textbf{k}}$. In this case, the same mechanism of intensive 
particle creation works also for any scalar fields including four  
Higgs bosons~\cite{Blaschke_04}
  \bea \label{cr-12} 
 \Omega_{\rm I~Higgs}=4\Omega_{g\rm I}. 
 \eea
The decays of the Higgs sector  including longitudinal vector $W$ 
and $Z$ bosons approximately  preserve this partial energy density 
for the decay products. These products  are   Cosmic Microwave 
Background (CMB) photons  and $n_{\nu}$ neutrino. 
Therefore, one obtains 
 \bea \label{cr-13} 
(1+n_{\nu})\Omega_{\rm CMB}\approx 4\Omega_{g\rm I}. 
 \eea

In our model there is the coincidence of two epochs: 
\begin{itemize}
\item the creation  
of SM bosons in the Universe in electroweak epoch
 \bea\label{c-5}1+z_{\rm W}= [M_W/H_0]^{1/3}=0.37\cdot10^{15},
 \eea
when the horizon $H(z_W)=(1+z_W)^2H_0=(1+z_W)^21.5\cdot 10^{-42}$GeV 
contains only a single $W$ boson; 
\item 
and the CMB origin time
 \bea\label{c-4}
 1+z_{\rm CMB} = [\lambda_{\rm CMB}H_0]^{-1/2}=[ 10^{-29}\cdot 2.35/1.5 ]^{1/2}=
0.39\cdot10^{15},
 \eea
 when the horizon contains only a single CMB photon with mean wave length $\lambda_{\rm CMB}$ that is  
approximately equal to the inverse temperature $\lambda^{-1}_{\rm 
CMB}=T_{\rm CMB}=2.35 \cdot 10^{-13}$~GeV.
 \end{itemize}
 In the same epoch 
$z_I\approx z_W \approx z_{\rm CMB}$, if the primordial graviton 
density~(\ref{cr-11}) coincides with the CMB density normalized to a 
single degree of freedom (as it was supposed in \cite{Gr-74}). The 
coincidence of the Planck epoch $z_I$  with the first two ones 
solves cosmological problems with the aid of the geometry of 
similarity (\ref{ratio-1}), without the inflation (see also 
\cite{Behnke_02}). 

While adding the SM sector to the theory in order to 
preserve the conformal symmetry, we should exclude the unique 
dimensional parameter from the SM Lagrangian, {\it i.e.} the Higgs 
term with a negative squared mass. However, following 
Kirzhnits~\cite{linde1}, we can include the vacuum 
expectation of the Higgs field (its zeroth harmonic) 
$\langle\phi\rangle$. 
The latter appears as a certain external initial data or a condensate.
In our construction we can choose it in the most simple form:
 $\langle\phi\rangle = Const = \langle\phi\rangle_I = 246$~GeV which 
could be consider as the initial condition at the beginning of the 
Universe. The fact, that the Higgs vacuum expectation is equal to 
its present day value, allows us to preserve the status of the SM as 
the proper quantum field theory during the whole Universe evolution. 
 The standard vacuum stability conditions 
 \bea
<0|0>|_{\phi=\langle \phi\rangle}=1, \qquad <0|0>'|_{\phi=\langle \phi\rangle}=0
 \eea
yield the following constraints on the Coleman--Weinberg effective potential
of the Higgs field:
 \bea 
 V_{\rm eff}(\langle \phi\rangle)=0, \qquad V'_{\rm eff}(\langle \phi\rangle)=0.
 \eea 
It results in a zero contribution of the Higgs field vacuum 
expectation into the Universe energy density. In other words, the SM 
mechanism of a mass generation can be completely repeated. However, 
the origin of the observed conformal symmetry breaking is not a 
dimensional parameter of the theory but a certain non-trivial (and 
very simple at the same moment) set of the initial data. In 
particular, one obtains that the Higgs boson mass is determined from 
the equation $V''_{\rm eff}(\langle \phi\rangle)=M_H^2$. Note that 
in our construction the Universe evolution is provided by the 
dilaton, without making use of any special potential and/or any 
inflaton field. In this case we have no reason to spoil the 
renormalizablity of the SM by introducing the non-minimal 
interaction between the Higgs boson and the gravity~\cite{Bezrukov:2007ep}.

In summary, following the ideas of the conformal symmetry 
\cite{Dirac_73,Weyl_18}, we formulated the GR in terms of the  
spin-connection coefficients. The cosmological evolution of the 
metrics is induced by the dilaton, without the inflation  hypothesis 
and the $\Lambda$-term.  In the suggested model, 
the Planck epoch coincides with the thermalization and the 
electroweak ones. In this case the CMB power spectrum can be 
explained by two gamma processes of SM bosons~\cite{2009}, avoiding 
dynamical dilaton deviations with negative energy by means of the 
Dirac constraint~(\ref{dirac-1}). We have provided a few arguments 
in favour  that the exact evolution of the GR as a theory of 
spontaneous  conformal symmetries breaking is related to the 
equations for the quantum squeezed oscillator. We found that the 
dilaton evolution yields the vacuum creation of matter.

\subsection*{Acknowledgments}
The authors thank    D.~Blaschke, K.~Bronnikov, D.V.~Gal'tsov, 
A.V.~Efremov, N.K.~Plakida, and V.B.~Priezzhev for useful 
discussions. V.N.P. thanks Yu.G.~Ignatev and N.I.~Kolosnitsyn 
for the discussion of experimental consequences of 
the General Relativity.


\begin{thebibliography}{99}

\bibitem{SN2}
                      A. G. Riess {\it et al.}, 
                      Astron. J. {\bf 116} (1998) 1009;
                      S. Perlmutter {\it et al.}, 
                      Astrophys. J. {\bf 517} (1999) 565;
                      P. Astier  {\it et al.}, 
                      Astronomy and Astrophysics {\bf 447} (2006) 31. 

\bibitem{eds32} 
                      A. Einstein and W. de-Sitter, 
                      Proc.  Nat. Acad. of Scien. {\bf 18} (1932) 213.

\bibitem{Linde:2007fr}
                      A. D. Linde,
                      Lect.\ Notes Phys.\  {\bf 738} (2008) 1.

\bibitem{gio}
                      M. Giovannini, 
                      Int. Jour. Mod. Phys. D{\bf 14} (2005) 363.
\bibitem{Grum}
                      D. Grumiller, W. Kummer, and D. V. Vassilevich,
                      Phys. Rep.  {\bf 369} (2002) 327.
 \bibitem{Dirac_73}
                      P. A. M. Dirac, 
                      Proc. Roy. Soc. A{\bf 333} (1973) 403.

\bibitem{Weyl_18}
                      H. Weyl, 
                      Sitzungsber. d. Berl. Akad., 465 (1918).

\bibitem{Behnke_02}
                      D. Behnke {\it et al.},
                      Phys. Lett. B{\bf 530} (2002) 20;
A.F.~Zakharov and V.N.~Pervushin,
  arXiv:1006.4745 [gr-qc].

\bibitem{Barbashov_06}
                      B. M. Barbashov {\it et al.}, 
                      Phys. Lett. B{\bf 633} (2006) 458; 
                      Int. Jour. Mod. Phys. A{\bf 21} (2006) 5957;
                      Int. J. Geom. Meth. Mod. Phys. {\bf 4} (2007) 171.

\bibitem{dir}
                      P. A. M. Dirac, Proc. Roy. Soc. A{\bf 246} (1958) 333;
                      Phys. Rev.{\bf 114} (1959) 924.

\bibitem{ADM}
                      R. Arnowitt, S. Deser, and C. W. Misner, 
                      {\sl The dynamics of general relativity}, 
                      in L. Witten, 
                      {\sl  Gravitation: An Introduction to Current Research} 
                      (Wiley, New York, 1962) pp.227-265.

\bibitem{lich}
                      A. Lichnerowicz, 
                      Journ. Math. Pures and Appl. {\bf B37} (1944) 23.

\bibitem{Misner_69}
                      C. Misner, 
                      Phys. Rev. {\bf 186} (1969) 1319.
\bibitem{Gr-74}
                      L. P. Grishchuk,
                      Sov. Phys. Usp. {\bf 20} (1977) 319.

\bibitem{ps1}
                      V.N. Pervushin and V.I. Smirichinski, 
                      J. Phys. A{\bf 32} (1999) 6191.

\bibitem{Fad}
                      L. D. Faddeev and V. N. Popov,
                      Sov. Phys. Usp.  {\bf 16} (1974) 777.

\bibitem{z-06}
                      A. F. Zakharov, V. A. Zinchuk, and V. N. Pervushin,
                      Phys. Part. Nucl. {\bf 37} (2006) 104.

\bibitem{grib88}
                      L. Parker, 
                      Phys. Rev. {\bf 183} (1969) 1057.
\bibitem{BR}  
                      J. P. Blaizot and G. Ripka, 
                     {\sl Quantum Theory of Finite Systems}
                     (The MIT Press, London, 1986).

\bibitem{sh-79}
                      J. Schwinger, L. DeRaad, and K. A. Milton,
                      Ann. Phys. {\bf 115} (1979) 1.

\bibitem{Blaschke_04}  
                      V.N. Pervushin,
                      Acta Phys. Slov. {\bf 53} (2003) 237;
                      D. B. Blaschke {\it et al.}, 
                      Phys. Atom. Nucl. {\bf 67} (2004) 1050.


\bibitem{linde1}
                      D. A. Kirzhnits, 
                      JETP Lett. {\bf 15} (1972) 529;
                      A. D. Linde, JETP Lett. {\bf  19} (1974) 183.

\bibitem{Bezrukov:2007ep}
                      F. L. Bezrukov and M. Shaposhnikov,
                      Phys. Lett. B{\bf 659} (2008) 703.

\bibitem{2009}
                      A. B  Arbuzov {\it et al.},
                      Physics of Atomic Nuclei {\bf 72} (2009) 744.
 
\end{thebibliography}
\end{document}